\documentclass[superscriptaddress,onecolumn,showpacs,showkeys,amsmath,preprint,aps]{revtex4}

\usepackage{graphicx}
\usepackage[sort&compress]{natbib}
\usepackage{ifthen}
\usepackage{hyperref}

\begin{document}
\title{Vortex pattern in a nanoscopic cylinder}
\author{Antonio R. de C. Romaguera}%
\affiliation{ Universidade Federal do Rio de Janeiro, Caixa
Postal:68.528, Rio de Janeiro - RJ, 21941-972,Brazil}%
\affiliation{Universiteit Antwerpen, Groenenborgerlaan 171, B-2020
Antwerpen, Belgium}
\author{Mauro M.Doria}%
\email{mmd@if.ufrj.br}\homepage{http://www.if.ufrj.br/~mmd}%
\affiliation{ Universidade Federal do Rio de Janeiro, Caixa
Postal:68.528, Rio de Janeiro - RJ, 21941-972,Brazil}%
\author{F. M. Peeters}%
\email{francois.peeters@ua.ac.be}\homepage{http://www.cmt.ua.ac.be}
\affiliation{Universiteit Antwerpen, Groenenborgerlaan 171, B-2020
Antwerpen, Belgium}
\date{\today}
\begin{abstract}
A superconducting nanoscopic cylinder, with radius $R = 4.0\xi$ and
height $D = 4.0\xi$ is submitted to an applied field along the
cylinder axis . The Ginzburg-Landau theory is solved in
three-dimensions using the simulated annealing technique to minimize
the free energy functional. We obtain different vortex patterns,
some of which are giant vortices and up to twelve vortices are able
to fit inside the cylinder.
\end{abstract}
\maketitle
Recently, a diode based on vortex physics was proposed\cite{SVMM06}.
It exhibits new features and opens new prospects for flux line
circuit implementations. Potential candidates for resistor,
capacitor and inductor based on vortex physics were also considered
and they all rely on the choice of sample geometry. Vortex patterns
have been previously studied in several geometries
with nanoscopic dimensions, such as disks\cite{SP98},
triangles\cite{MBM04} and squares\cite{CCBM00}. They were
theoretically studied using the Ginzburg-Landau (G-L) theory but
only in the two-dimensional limit, where the sample thickness is
smaller than $\xi$. This allowed them to assume that the order
parameter is constant along the third direction defined by the field
reducing the dimensionality of the problem. The other limit of an
infinite long cylinder was studied in \cite{DZ02}. Here we consider
the intermediate case of a nanoscopic cylinder with radius
$R=4.0\xi$ and height $D=4.0\xi$ using the full three-dimensional
approach and obtain the vortex states as a function of the applied
field perpendicular to its surface.

\begin{figure}[t]
\centering
\includegraphics[width=\linewidth]{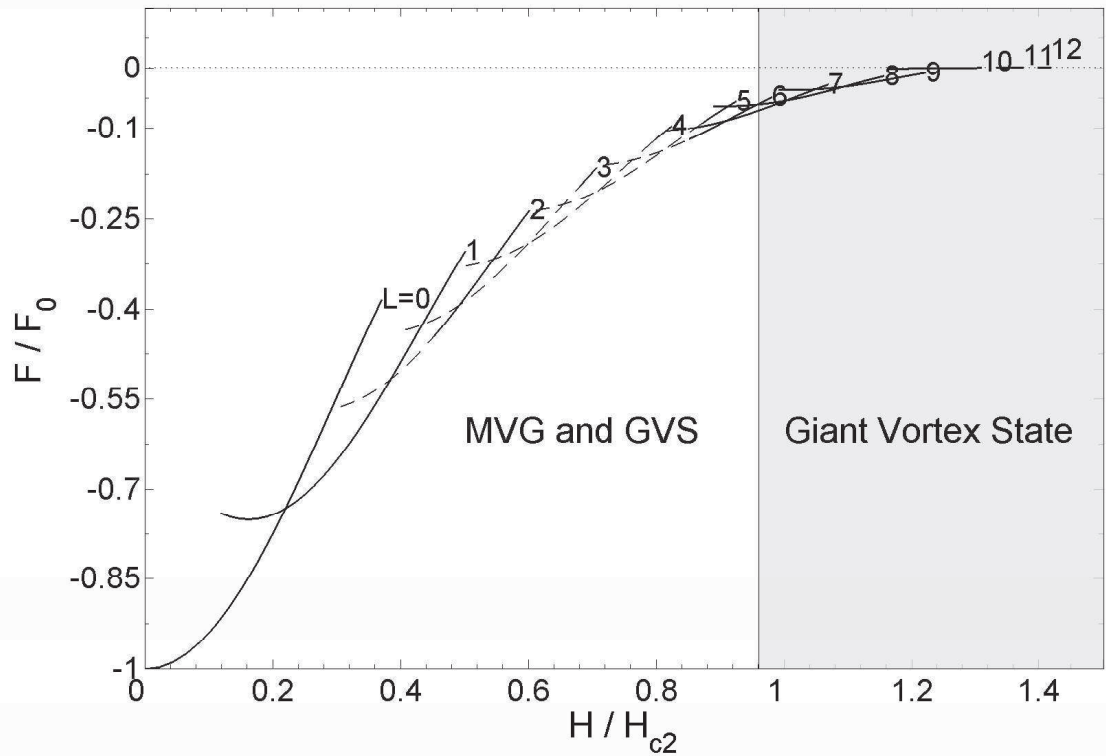}
\caption{Free energy of the GVS (solid curves) and the MVS (dashed
curves) versus  field for a cylinder of radius $R=4.0\xi$ and
thickness $D=4.0\xi$. The labels inside the curves indicate the
vorticity of the system. In the grey region only GVS were observed.
In the white region both MVS and GVS were observed.}\label{fig1}
\end{figure}

\begin{figure}[t]
\centering
\includegraphics[width=\linewidth]{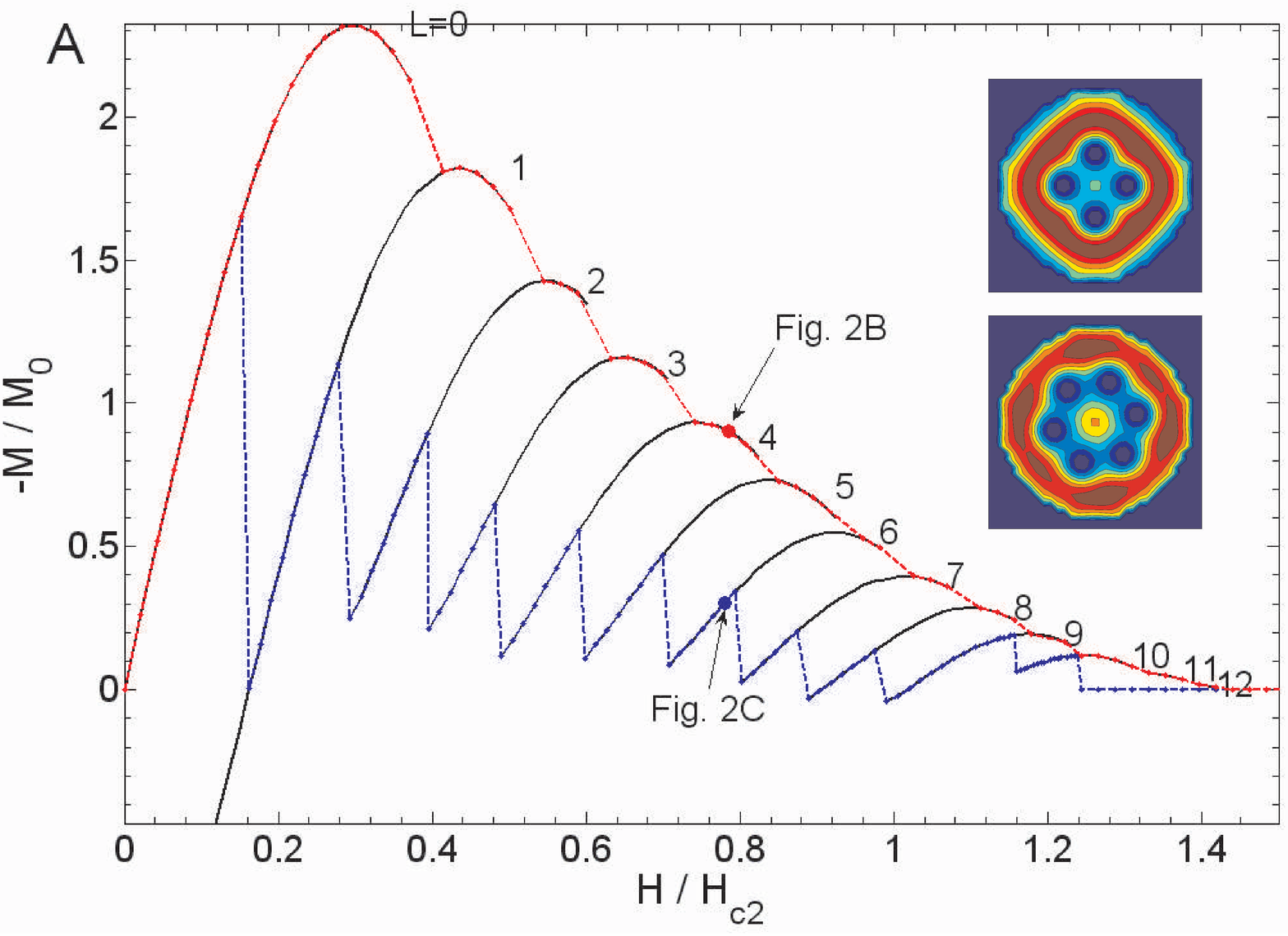}
\includegraphics[width=0.45\linewidth]{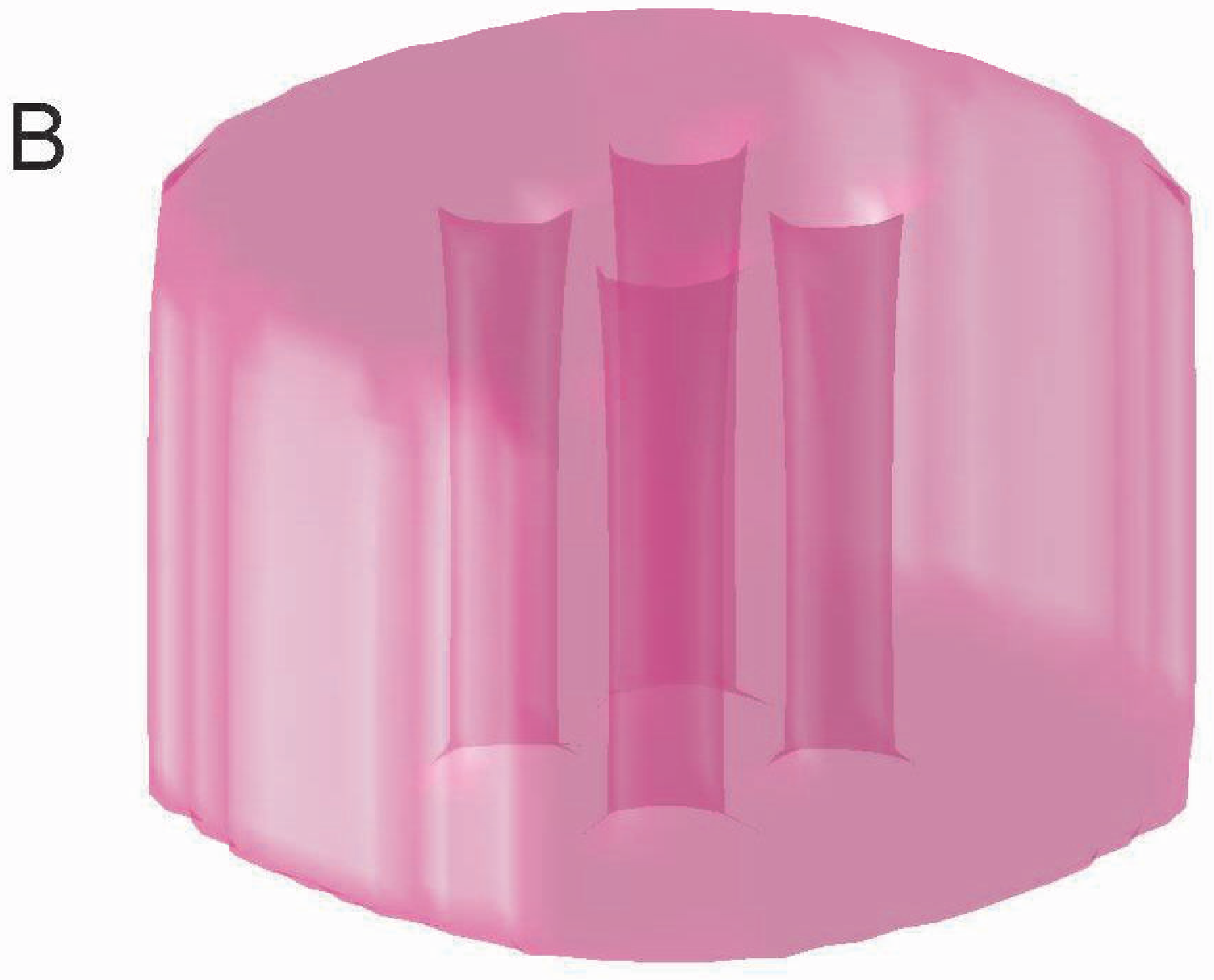}
\includegraphics[width=0.45\linewidth]{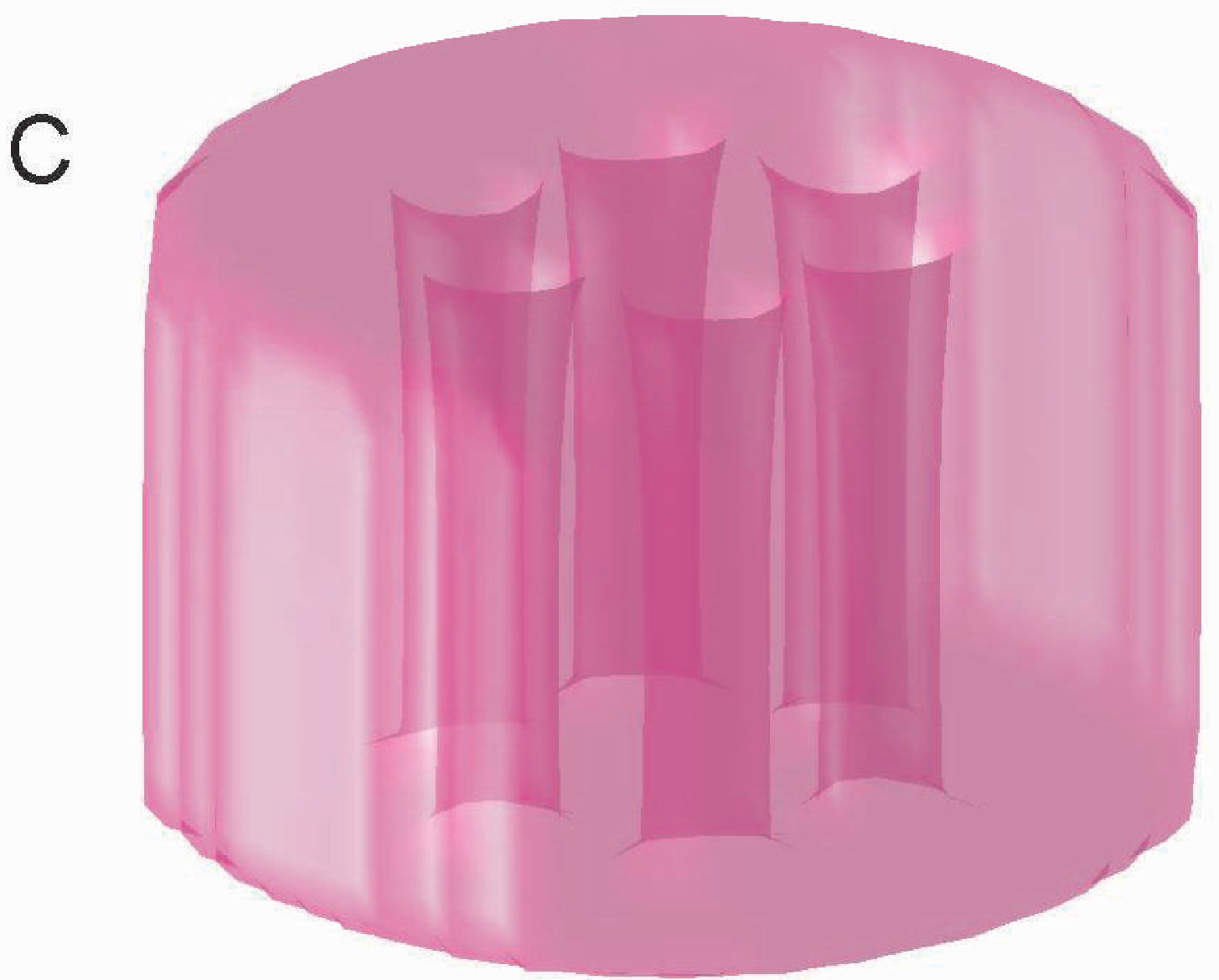}
\caption{\textbf{(A)} Magnetization versus applied field for the
cylinder. The insets are the two dimensional $|\Psi(\vec{r})|^2=cte$
plots for $H/H_{c2}=0.785$ with vorticity 4 and 6. \textbf{(B)} and
\textbf{(C)} show the three dimension Cooper pair density
isosurface.}
\label{fig2}%
\end{figure}

We minimize numerically the G-L free energy functional using the
simulated annealing procedure, discussed in Ref. \cite{RD04} and in
references quoted therein. Every vortex state is characterized by an
integer number which is the total vorticity inside the cylinder. In
some magnetic field range we find for increasing field and constant
vorticity that the vortex evolves from a set of individual vortices
with unitary vorticity to another set containing vortices with
vorticity larger than one, the so called giant vortex. However for
the same conditions of increasing field and constant vorticity the
pattern becomes metastable as another vortex state becomes lower in
free energy. This change of stability takes place at the matching
field\cite{SDN99}. In figure \ref{fig1} we show the free energy
versus the applied field and in figure \ref{fig2} the corresponding
magnetization.
The lowest energy curve corresponds to the thermodynamic most stable
pattern, the Meissner phase with no vortices, but only up to the
first matching field. Above this field the entrance of a single
vortex is favored, though it is still possible to keep the system in
the Meissner phase but now as a metastable configuration. The small
difference in energy around the matching field is able to drive the
state to a magnetic hysteresis. In figure \ref{fig1} we find Multi
Vortex States (MVS) and Giant Vortex States (GVS) for fields below a
threshold value $H/H_{c2} = 0.96$ and only GVS above it. This result
is different from those in Ref \cite{BP02} where they found only GVS
as ground states (MVS appeared as metastable states). The red and
blue lines in the magnetization, figure \ref{fig2}, present the full
sweep field up and down, respectively. The magnetization curve is
truly a collection of several independent lines, each associated to
a distinct vorticity which do not cross each other. As the applied
field is swept the vortex pattern jumps from one curve to the next
one causing the vorticity to jump with one unit and leading to
hysteretic behavior. Figures 2B and 3C show the Cooper pair density
of two different states corresponding to the field $H/H_{c2}=0.785$.
The state with $L=4$ was obtained with increasing the field(red
curve) and the state $L=6$ with decreasing the field. We find twelve
independent magnetization lines which implies that a total of twelve
vortices fit inside the cylinder. This result is in fair agreement
with the eleven vortices found by Baelus et al. \cite{BP02} for a
disk with the same radius, thickness $0.1\xi$ and $\kappa=0.28$.
Their study is based on a different method, the time evolution of
the corresponding two-dimensional system. The maximum number of
vortices inside a cylinder follows from the ratio between two areas,
the disk surface and the vortex core: $n_s= R^2/\xi^2=16$. This
value overestimates the true result found here because it does not
consider the repulsion among vortices. The first vortex appear at
the field $H/H_{c2}\cong 0.21$ and the normal phase is reached at
the applied field $H_{c3}/H_{c2}\cong 1.48$. For a disk \cite{BP02}
larger values were found for these critical fields, namely, 0.38 and
1.95 in units of $H_{c2}$ (which is a consequence of the smaller
expulsion of the magnetic field in a thin disk).

\textbf{Acknowledgment} \label{labelOfAcknowledgment} A. R. de C.
Romaguera and M. M. Doria thank the Brazilian agencies CNPq, FAPERJ
and the Instituto do Mil\^enio de Nanotecnologia for financial
support. F. M. Peeters acknowledges support from the Flemish Science
Foundation (FWO-Vl), the Belgian Science Policy (IUAP), the
JSPS/ESF-NES program and the ESF-AQDJJ network.
\bibliographystyle{phjcp}
\bibliography{RDPm2sv6}
\end{document}